\documentclass[a4paper,aps,preprintnumbers,twocolumn,amsmath,amssymb,prl,showpacs]{revtex4}
\usepackage{bbm}
\usepackage{mathrsfs}
\usepackage{amsmath}
\usepackage{graphicx}
\usepackage{color}

\newcommand{\be}{\begin{equation}}
\newcommand{\ee}{\end{equation}}
\newcommand{\bea}{\begin{eqnarray}}
\newcommand{\eea}{\end{eqnarray}}
\newcommand{\bes}{\begin{split}}
\newcommand{\ees}{\end{split}}

\renewcommand{\vec}[1]{\mathbf{#1}}

\newcommand{\T}{\operatorname{T}}

\newcommand{\tr}{\operatorname{Tr}}

\newcommand{\mb}{\mbox}

\begin{document}
\title{Quench Dynamics of  Topological  Maximally-Entangled  States}
\author{Ming-Chiang Chung$^{1}$, Yi-Hao Jhu$^2$, Pochung Chen$^2$, and Chung-Yu Mou$^{1,2,3}$ }
\affiliation{$^1$Physics Division, National Center for Theoretical Science, Hsinchu, 30013, Taiwan}
\affiliation{$^2$Physics Department, National Tsing Hua University, Hsinchu, 30013, Taiwan}
\affiliation{$^3$Institute of Physics, Academia Sinica, Taipei 11529, Taiwan}

\begin{abstract}
We investigate the quench dynamics of the one-particle entanglement spectra (OPES) for systems with 
topologically nontrivial phases. By using dimerized chains as an example, it is demonstrated that the evolution 
of OPES for the quenched bi-partite systems is governed by an effective Hamiltonian which is characterized 
by a pseudo spin in a time-dependent pseudo magnetic field $\vec{S}(k,t)$. The existence and evolution of 
the topological maximally-entangled edge states are determined by the winding number of $\vec{S}(k,t)$ 
in the $k$-space. In particular, the maximally-entangled edge states survive only if nontrivial 
Berry phases are induced by the winding of $\vec{S}(k,t)$. In the infinite time limit the 
equilibrium OPES can be determined by an effective time-independent pseudo magnetic field $\vec{S}_{\mb{eff}}(k)$.
Furthermore, when maximally-entangled edge states are unstable, they are destroyed by quasiparticles within 
a characteristic timescale in proportional to the system size.
\end{abstract}

\pacs{73.20.At, 03.67.Mn,05.70.Ln}

\date{\today}
\maketitle

Topological phases that are not  characterized by local order parameters have been subjects of 
key interests in condensed matter physics due to anomalous properties associated with these phases.
Recent discovery  of time-reversal invariant  topological insulators \cite{Topo,TopoExp} has further
triggered intense investigations on the characterization of topological phases. 
One of the important features for topological phases is the possibility of creating
nonlocal properties associated with the topology, which is often realized as the entanglement between
the system and its environment. In particular, the entanglement spectrum (ES) i.e. the eigenvalues
of the reduced density matrix of the system, provides an idea tool to characterize the 
topological phase \cite{Haldane, KitaevPollmann}. In general, not only the ES can be used to distinguish 
different classification of topological phases but also it can  detect the existence of edge modes 
at zero energy \cite{RyuHatsugai02, MJCY11}.  The existence of edge states at zero energy reflects 
the non-trivial topology of the underlying quantum state. From the point of view for manipulating 
quantum information, these edge modes represent the maximally entangled  states of the
system and its environment. Therefore, they are the best candidate for qubits \cite{MJCY11}. 

In order for the topological  maximally-entangled state being a viable candidate for qubits, 
it is necessary to examine if they could survive under quantum information
processing. Since typical quantum manipulations involve rapid change of the coupling to the environments, 
it is therefore important to examine the quench dynamics of maximally-entangled  states. 
Recent investigations indicate that the thermalization of integrable systems due to quench depends 
strongly on the initial conditions \cite{ColdAtoms,Rigol, MingMiguelAnibal}. These studies, however, 
are confined to the bulk properties. There are only few papers
concerning quench dynamics of topological edge states
\cite{Delgado}. In the presence of topological  edge states, the system can be maximally  
entangled with the environment and the thermalization of quench dynamics could be entirely different. 
It is thus important to examine the quench dynamics of  topological  maximally-entangled  states. 
In this paper, by taking dimerized chain as an example, we investigate how edge states affect 
the thermalization and the quench dynamics of one-particle
entanglement spectra (OPES)  defined below. 
In particular, we show that the existence 
and evolution of topological maximally-entangled edge states are determined by 
the winding number of a pseudo magnetic field $\vec{S}(k,t)$. The maximally-entangled states survive
only if nontrivial Berry phases are induced by the winding of $\vec{S}(k,t)$.

Consider the ground state $|\Psi_{AB}\rangle$ of a bipartite total system $AB$ that consists of the system $A$ 
and the environment $B$. The reduced density matrix of the system $A$ is $\rho_A=\tr_B |\Psi_{AB}\rangle\langle\Psi_{AB}|$.
The entanglement entropy (EE), defined as $S_A \equiv -\tr \rho_A \log_{2} \rho_A$, 
has been widely used to measure the bipartite entanglement between the system $A$ and the environment $B$ \cite{Review}.
It is known that the scaling law of EE provides a way to distinguish different quantum phases \cite{AreaLaw}.
Furthermore, the property that EE diverges at the critical
points provides a useful tool to examine the quantum criticality \cite{QCP}. 
In addition to global properties associated with EE, it is useful to explore detailed microscopic quantum phenomena
using OPES, defined as the set of $\lambda_m$'s with
$\rho_A = \bigotimes_{m} \left[\begin{matrix}  \lambda_m & 0\\ 0 & 1-\lambda_m\end{matrix} \right]$.
EE and OPES are related through the relation $S_A =\sum_m S_m$ where
$\label{Sent} S_m = -  \lambda_m \log_2{\lambda_m} - (1-\lambda_m)  \log_2{(1-\lambda_m)}$.
The OPES has been used to investigate disorder lines \cite{MingPeschel}, Berry phase \cite{RyuHatsugai06, MJCY11} 
and zero-energy edge states \cite{RyuHatsugai06, MJCY11}. 
It is clear that the eigenvalue $\lambda_m=1/2$ corresponds to the situation when the system $A$ and 
the environment $B$ are maximally entangled so that $S_m=1$. Since $\lambda_m=1/2$ is a mid-gap state that often results from
the zero-energy edge state between $A$ and $B$, the maximally-entangle state is often topologically protected.

To investigate quench dynamics of the maximally-entangled state, we
consider a one-dimensional (1D) dimerized  chain 
characterized by the Hamiltonian:
\be 
  \label{eq:DimerH}
  H_\delta = -\sum_i (1+\delta (-1)^{i}) (c_i^{\dagger} c_{i+1} + {\mbox{h.c.}}),
\ee
where $i$ is the site index and $\delta\in [-1,1]$. 
The ground state undergoes a phase transition from a topologically trivial 
phase ($\delta < 0$) to a topologically nontrivial phase  ($\delta > 0$) as $\delta$ is
varied across the phase boundary $\delta_c=0$. Let the region $i >0 $ represents the system $A$ 
while $i \leq 0$ represents the environment $B$, the topological edge mode appears if the bond 
$[i,i+1] = [0,1]$ is strong, i.e., $-(1+\delta) < - (1-\delta)$. 
Otherwise the edge mode does not exist \cite{wu}. 
In this paper we study the quench dynamics of the topological
maximally-entangled states for a dimerized chain by suddenly
quenching the  parameter $\delta$. 

\begin{figure}
\center
\includegraphics[width=7.5cm]{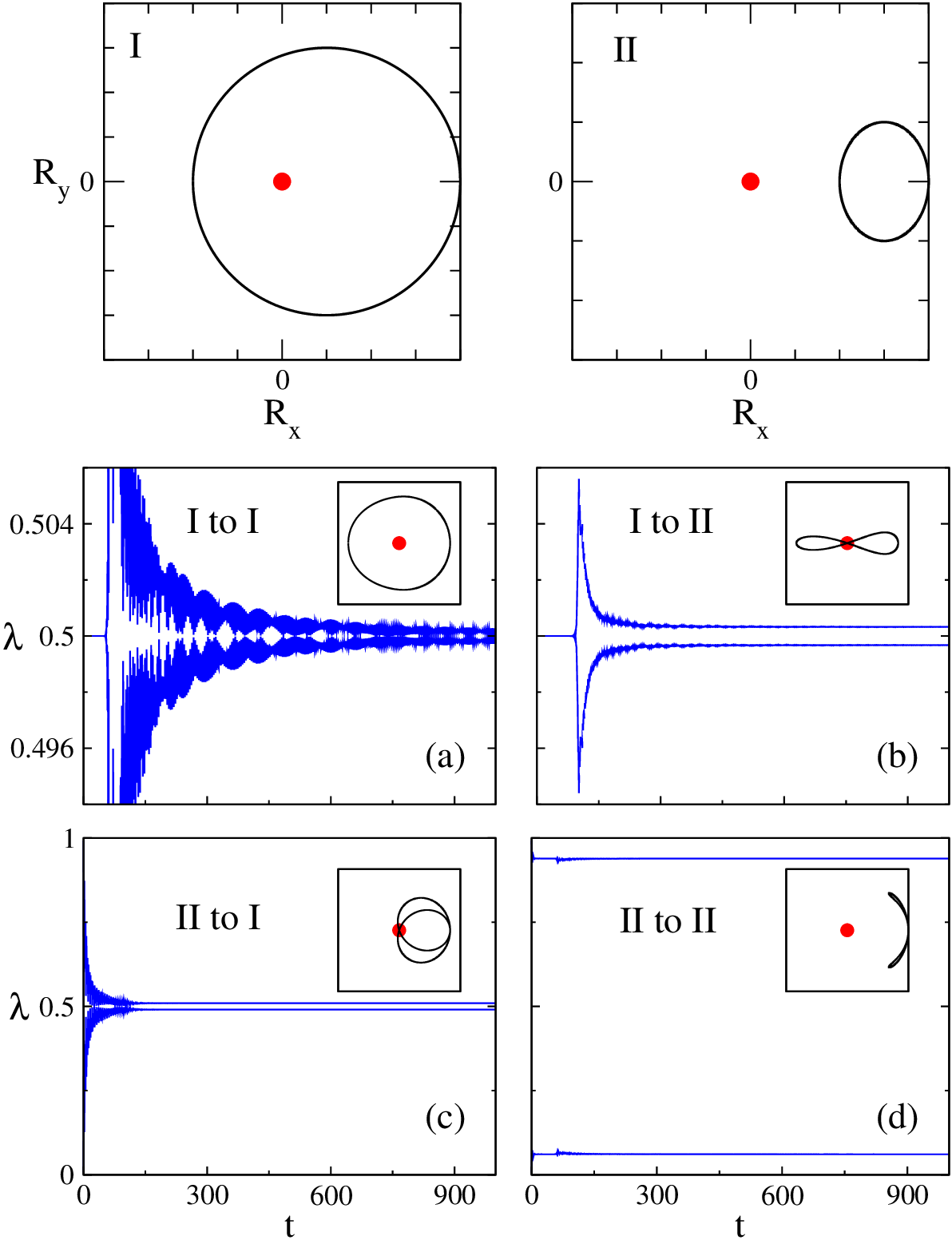}
\caption{(color online) 
Upper two panels: 
Representative loops $\ell$ of $\vec{R}_\delta(k)$ for phase I and II 
with $\delta=0.5$ and $\delta=-0.5$ respectively.
In phase I ($\delta \in (0,1]$) the Berry phase is $\pi$ due to circling of $\vec{R}_\delta(k)$ 
around the origin, while in phase II ($\delta \in [-1,0)$) the Berry phase is trivial. 
Lower four panels: evolutions of two $\lambda_m(t)$ that are closest to $1/2$ 
for different quench processes: 
(a) phase I ($\delta=0.5$) to I ($\delta=0.2$) 
(b) phase I ($\delta=0.5$) to II ($\delta=-0.5$) 
(c) phase II ($\delta=-0.5$) to I ($\delta=0.5$) 
(d) phase II ($\delta=-0.5$) to II ($\delta=-0.2$). 
Insets: loop $\ell$ of $\vec{S_{\mb{eff}}}$}.
\label{fig1}
\end{figure}

The occurrence of maximally-entangled edge states has its topological origin.  
By defining a spinor $\vec{c}_{x_i} = (c_{x_{i1}},c_{x_{i2}})^{\T}$ where $x_{i1}=2i-1$ and $x_{i2}=2i$
and performing a Fourier transformation, the Hamiltonian (\ref{eq:DimerH}) can be casted into the form
\be
  \label{Htopo}
  H_\delta = - \sum_{k \in BZ} \vec{c}_{k}^{\dagger} [\vec{R}_\delta(k)\cdot \boldsymbol{\sigma} ]\vec{c}_{k},
\ee 
where $\boldsymbol{\sigma} = (\sigma_x,\sigma_y,\sigma_z) $ are Pauli matrices and 
\be
  \vec{R}_\delta(k) = \left((1- \delta) + (1 + \delta) \cos{k},(1 + \delta) \sin{k},0 \right) 
\ee
is a pseudo magnetic field with the magnitude $R_\delta(k) = 2 (\cos^2{k/2}+\delta^2 \sin^2{k/2})^{1/2}$. 
The system fulfills chiral symmetry because $\vec{R}_\delta(k)$ lies on a plane. For $\delta \in (0,1]$ 
the loop $\ell$ of $\vec{R}_\delta(k)$ encloses the origin $\cal{O}$  as $k$ runs through the Brillouin zone.
Consequently, the Berry phase (or Zak's phase) $\chi$, defined as a line integral of the curvature of the
filled band, is $\pi$. 
Due to the fact that $\ell$ can be  continuously  deformed
into a unit circle without crossing the origin,  topological argument ensures that the original Hamiltonian
corresponding to $\ell$ contains at least one pair of one zero-energy edge states
as a consequence of chiral symmetry \cite{RyuHatsugai02}.  On the other
hand, if $\delta \in [-1,0)$,  one obtains trivial Berry phases and
no topological edge state occurs (see the upper panels of Fig~\ref{fig1}).  

We consider a total system of infinite size, and partition it into the system $A$ 
with $x_i=1,\cdots, L/2$ and its environment $B$ (the remaining part).
The reduced density matrix of the system $\rho_A$ for the ground state
can be determined  by eigenvalues $\lambda_m$ of the block Green's function matrix (GFM) \cite{GFM} 
$G^\delta_{\alpha\beta}(x_i, x_j) = \tr [\rho c_{x_{i\alpha}}^{\dagger} c_{x_{j\beta}}]$, 
where $x_{i\alpha}, x_{j\beta}$ belong to the system $A$, $\alpha,\beta=1,2$,
and $\rho$ is the ground state density matrix of the total system.
Hence the GFM can be considered as an effective Hamiltonian that
determines the OPES. 
In the Fourier $k$-space, $G^\delta_{\alpha\beta}(k)$ is given by
\be \label{GTopo}
    G^\delta_{\alpha\beta}(k) = \frac{1}{2}  \left[ 1 + \hat{\vec{R}}_\delta(k) \cdot \boldsymbol{\sigma} \right]_{\alpha\beta},
\ee
where $\hat{\vec{R}}_\delta(k)=\vec{R}_\delta(k)/R_\delta(k)$.
It takes almost the same form as the Hamiltonian (\ref{Htopo}) except for a constant and a positive normalization
factor $R_\delta$, which leads to the conclusion that they share the same topology. If $\vec{R}_\delta(k)$ encloses the origin 
in the parameter space with Berry phase equal to $\pi$, a pair of zero energy states appear for the Hamiltonian (\ref{Htopo}), 
while for the GFM, we obtain a pair of  maximally-entangled states with $\lambda_m =1/2$. Notice that the maximally-entangled states 
are also edge states, which can be seen in Fig~\ref{fig3}. 

Consider now a sudden quench at $t=0$ by changing the parameter of the dimerized chain 
in Eq.~(\ref{eq:DimerH}) from $\delta$ at $t<0$ to $\delta^\prime$ at $t>0$.
Denote the phase whose $\delta \in (0.1]$ with maximally-entangled states  by phase I,
and the phase whose $\delta \in[-1,0)$ by phase II. 
We perform four possible quench processes, including I to I, I to II, II to I, and II to II. 
The OPES $\lambda_m(t)$ at time $t$ can be obtained by diagonalizing 
a time-dependent GFM which is defined as 
$G_{\alpha\beta}(x_i,x_j,t)
=\tr (\rho e^{i H_{\delta^\prime}t} c_{x_{i\alpha}}^{\dagger} c_{x_{j\beta}} e^{-iH_{\delta^\prime}t} )$
where  $x_i, x_j \in A$ \cite{CalabreseCardy}.  
In Fig~\ref{fig1} (a)-(d), we show the time evolution of OPES for states whose eigenvalues are closest to $1/2$.  
In cases of I to I and I to II, 
the maximally-entangled states persist for a while and then split into two different evolutions. 
Only for the case of I to I, however, the two splitting eigenstates evolve back to 
two maximally-entangled states at infinite time.
In contrast, for the case of I to II, two splitting eigenstates closest to $1/2$ remain splitting forever. 
On the other hand, if one starts with initial states without maximally-entangled states such as the cases of 
II to I and II to II, maximally-entangled states can not be created at later time.

It is instructive to define a time-dependent pseudo magnetic field $\vec{S}(k,t)$ 
from the time-dependent GFM $G(k,t)$ in the Fourier space through the relation
\be
  G_{\alpha\beta}(k,t) = 1/2 \left[ 1 + \vec{S}(k,t) \cdot \boldsymbol{\sigma} \right]_{\alpha\beta}.
\ee
$\vec{S}(k,t)$ can be further written as a summation of three different contributions 
\bea
  \label{G1} \vec{S}_1(k,t) & = & \cos(2R_{\delta^\prime}t) \hat{\vec{R}}_\delta(k), \\
  \label{G2} \vec{S}_2(k,t) & = & \sin(2R_{\delta^\prime}t) 
  \hat{\vec{R}}_{\delta}(k) \times \hat{\vec{R}}_{\delta^\prime}(k), \\
  \label{G3} \vec{S}_3(k,t) & = & (1 - \cos(2R_{\delta^\prime}t)) 
  \hat{\vec{R}}_\delta(k) \cdot \hat{\vec{R}}_{\delta^\prime}(k) \hat{\vec{R}}_{\delta^\prime}(k). 
\eea
The thermalization of EE observed in Fig~\ref{fig1} can be understood by considering the infinite-time limit 
of the time-dependent pseudo magnetic field (\ref{G1}), (\ref{G2}), and (\ref{G3}). 
Since the sinusoidal parts dephase out, the  long-time behavior of
the time-dependent GFM, $G_{\alpha\beta}(k,t=\infty)$, is solely determined by the effective pseudo magnetic field 
\be \label{Reff}
  \vec{S}_{\mb{eff}}(k) \equiv 
  \left(\hat{\vec{R}}_\delta(k) \cdot \hat{\vec{R}}_{\delta^\prime}(k) \right)\hat{\vec{R}}_{\delta^\prime}(k). 
\ee 
The existence of  maximally-entangled states at infinite time is thus
determined by the topology of $\vec{S}_{\mb{eff}}$. If $\vec{S}_{\mb{eff}}$
encircles origin, the Berry phase is $\pi$, maximally-entangled
states appear at $t = \infty$; otherwise there is no maximally-entangled state at later time. 
It is clear that both the final $\vec{R}_{\delta^\prime}$ and the initial $\vec{R}_\delta$ 
determine $\vec{S}_{\mb{eff}}$. It implies that the long time behavior 
carries the memory of the initial state, which is due to nonergodicity of integrable
systems \cite{MingMiguelAnibal}. The existence of maximally-entangled states (edge states) 
at infinite time requires that both the initial and final
Hamiltonians possess nontrivial Berry phase. 
In the inset of Fig~\ref{fig1}(a)-(d), $\vec{S}_{\mb{eff}}$ are plotted for different quench processes
with different initial and final Hamiltonians. It is clear that only
for the quench from I to I,
$\vec{S}_{\mb{eff}}$ encircles origin, while in the other quenches $\vec{S}_{\mb{eff}}$ either not encircles the
origin (II to II) or passes through the origin (I to II or
II to I). 

\begin{figure}
\center
\includegraphics[width=7.5cm]{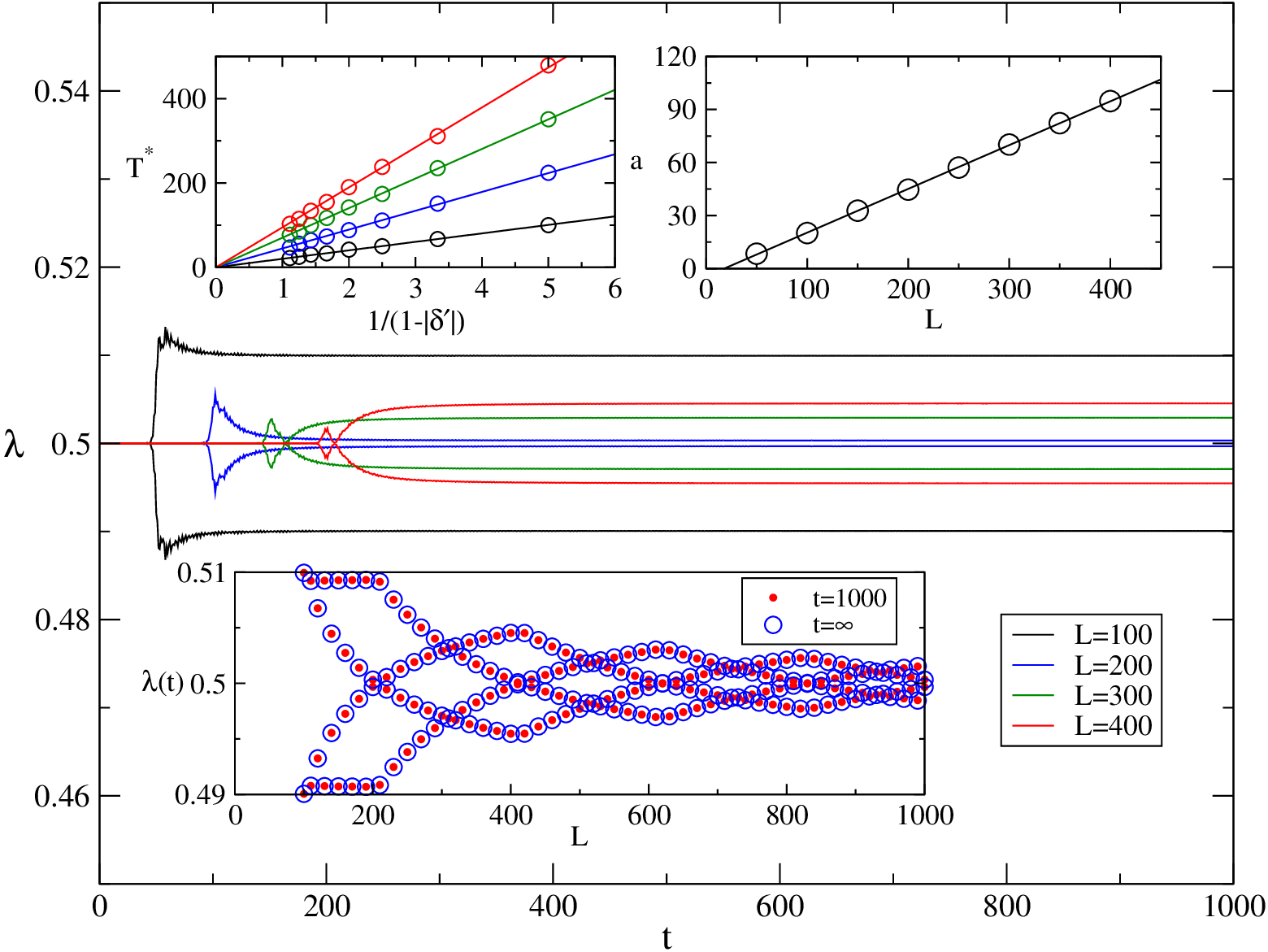}
\caption{(color online)
Evolutions of two $\lambda_m(t)$ that are closest to $1/2$ for different $L$
under the quench of I ($\delta=0.5$) to II ($\delta=-0.5$).
Lower inset: $\lambda_m(t)$ obtained from exact diagonalization of 
$G_{\alpha,\beta}(x_i,x_j,t=1000)$ (blue circle) and $\vec{S_{\mb{eff}}}$ (red dots). 
Upper left panel: fitting of $T^{\star}$, the persistence time of the maximum entangled states, 
using $T^{\star}(L) = a(L) / (1-|\delta'|)$ where $T^{\star}$ is the persistence time of the maximum entangled states. 
Upper right panel: fitting of $a(L)$ versus $L$. We find that $a \approx 0.232L$.} \label{fig2}
\end{figure} 

The reason is that if the initial state is in the same phase as the final Hamiltonian 
(I to I or II to II), $\hat{\vec{R}}_\delta(k) \cdot \hat{\vec{R}}_{\delta^\prime}(k)$ 
is always positive, therefore the topology of $\vec{S}_{\mb{eff}}$ is the same as $\hat{\vec{R}}_{\delta^\prime}(k)$. 
On the other hand, if the system is quenched into a phase that is distinctly from 
the initial state, $\hat{\vec{R}}_\delta(k=0)\cdot\hat{\vec{R}}_{\delta^\prime}(k=0)$ is positive while
$\hat{\vec{R}}_\delta(k=\pi)\cdot\hat{\vec{R}}_{\delta^\prime}(k=\pi)$ is negative, 
then there must exist one point $k_0$ that $\vec{S}_{\mb{eff}}(k_0)  \propto
\hat{\vec{R}}_\delta(k_0)\cdot\hat{\vec{R}}_{\delta^\prime}(k_0) = 0$.
Hence, $k_0$ destroys the topology of the infinite-time Green's function matrix.  There exist
no maximally-entangled states whenever the system is quenched across
the topological phase boundary. This explains why only the quench from
phase I to phase I creates maximally-entangled states at infinite time. 

For all cases shown in Fig~\ref{fig1}, the OPES $\lambda_m(t)$ of the
dimerized chain fluctuate before it reaches the equilibrium. The intermediate regions can be explained by the appearance of  
$\vec{S}_2$ in the time-dependent GFM. Since $\vec{S}_2$ is proportional to 
$\hat{\vec{R}}_\delta(k) \times \hat{\vec{R}}_{\delta^\prime}(k)$ and is perpendicular to 
$\vec{S}_1$ and $\vec{S}_3$, the system is agitated by quasiparticles induced by $\vec{S}_2$ and $\vec{S}_3$ 
until the time-dependent sinusoidal functions dephase out and then the system reaches its equilibrate state 
with the effective $\vec{S}_{\mb{eff}}$  (\ref{Reff}). Clearly, the thermalization depends on sizes of the
system $A$. To check the size dependence, Fig~\ref{fig2} shows several quench processes 
from  phase I to phase II with different system length. In the
lower panel, we compare eigenvalues at $t=1000$ (red dots) evolved from the maximally-entangled state with eigenvalues obtained by 
diagonalization of $G_{\alpha\delta}(\vec{k},\infty)$ (blue circles). The dependence
on the system size $L$ shows excellent agreement, indicating the validity of $\vec{S}_{\mb{eff}}$.

We  also find that if the initially the system is in phase I, the maximally-entangled states can persist 
for a long time $T^{\star}$ before they are destroyed. This feature is striking because the
maximally-entangled states are not destroyed in the beginning. Furthermore $T^{\star}$ is independent of 
the initial conditions but depends on the final Hamiltonian and the system length $L$.
This is because the maximally-entangled states reside on the edge \cite{MJCY11}
and they will disappear only when the quasiparticles created from the
bulk  reach the edges. 
Therefore, we expect $T^{\star} \approx L/(2 v_{\mb{max}})$, where $v_{\mb{max}}$ 
is the maximum velocity of the quasiparticles. For the energy dispersion on obtains
$v_{\mb{max}} = 2(1-|\delta'|)$ hence
\be \label{Tstar}
 T^{\star} \approx \frac{L}{4 (1-|\delta'|)}. 
\ee
In the upper left inset of Fig~\ref{fig2}, we show the fitting of 
$T^{\star}(L)$ to the function $a(L)/(1-|\delta'|)$ using various system size $L$. 
We then fit $a(L)=b \times L$ as shown in the upper right inset of
Fig~\ref{fig2} to find $b\approx 0.232$. 
Combining these two fitting we find that $T^{\star} \approx 0.232 L/ (1-|\delta'|)$ which is very closed to
our approximation Eq.~(\ref{Tstar}). 
Our results imply that edge modes remains
to the infinite time if $L$ is infinite. Therefore, the edge modes
serve as good qubits since they have the maximal entanglements and 
cannot be destroyed easily if the system is large enough.      

\begin{figure}
\center
\includegraphics[width=7.5cm]{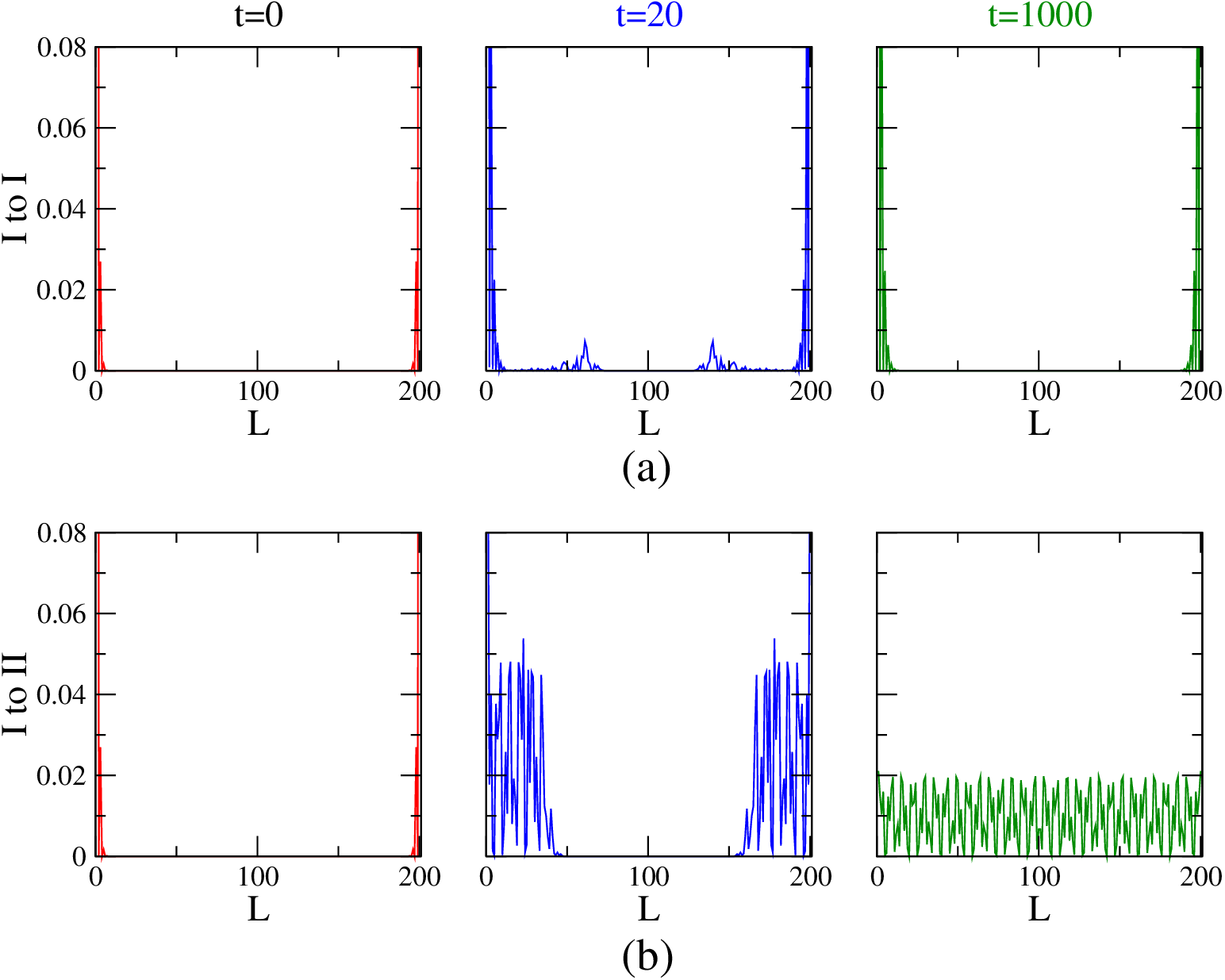}
\caption{The probability sum of the two eigenstates whose eigenvalues
are closest to $1/2$. (a) Quench process I to I. The edge
modes persist from the beginning to the end.  (b) Quench
process I to II. The edge modes disappear and evolve to the
bulk state.} \label{fig3}
\end{figure} 

The disappearance and recreation for different quench process starting from the phase I 
can be further explored by examining the
probability sum of the two eigenstates with
eigenvalues being closest to $1/2$. Fig~\ref{fig3}(a) shows the case of I to I. It
is seen that only small amounts of quasiparticles are excited. As a result, the edge
modes remain from the beginning to the end. On the other hand, for the quench process I to II as shown in Fig~\ref{fig3}(b),  quasiparticles strongly modify two edge modes until they disappear
and the whole system is bulk-like without any edge mode.

In summary, using dimerized chains as an example, the quench dynamics of the maximally-entangled states 
is investigated by diagonalizing the
time-dependent Green's function matrix.   We find that the existence of the maximally-entangled
states after sudden quench is determined by an effective pseudo magnetic field
($\vec{S}_{\mb{eff}}$), which depends on both the initial and final
Hamiltonians. The topological properties at 
infinite time are thus determined by the initial states and the
Hamiltonian after the quench. When the maximally-entangled states are
unstable, they are destroyed by quasiparticles that move from the
bulk to the edges with a characteristic timescale  proportional to the system size.

We thank Dr. Sungkit Yip for useful discussions. This work was
supported by the National Science Council of Taiwan.

\end{document}